\documentclass[12pt]{article}
\usepackage[T1]{fontenc} 
\usepackage[utf8]{inputenx}
\usepackage{babel} 
\usepackage{url}
\usepackage{geometry}
\usepackage{amsmath}
\usepackage{amsfonts}
\usepackage{amstext}
\usepackage{amssymb}
\usepackage{amsthm}
\usepackage{amscd}
\usepackage{mathrsfs}
\usepackage{xcolor}
\definecolor{myurlcolor}{rgb}{0,0,0.4}
\definecolor{mycitecolor}{rgb}{0,0.5,0}
\definecolor{myrefcolor}{rgb}{0.5,0,0}
\usepackage{graphicx}
\usepackage{tikz}
\usepackage{tikz-cd}
\usepackage{etaremune}
\usepackage{latexsym}
\usepackage{braket}
\usepackage{soul}
\usepackage{comment}

\usepackage{etoolbox}
\usepackage{makeidx}
\usepackage{dsfont}
\usepackage{enumitem} 
\usepackage{caption}
\usepackage{lmodern}
\usepackage{textcomp}
\usepackage{totcount}
\usepackage{blindtext}

\usepackage[pagebackref,draft=false]{hyperref}
\hypersetup{colorlinks,linkcolor=myrefcolor,citecolor=mycitecolor,urlcolor=myurlcolor}

\usepackage[capitalize]{cleveref}

\newtheorem*{proof*}{Proof}

\definecolor{darkpink}{rgb}{0.91, 0.33, 0.5}

\title{On the categorical foundations of quantum information theory: Categories and the Cramer-Rao inequality}

\author{F. M. Ciaglia$^{1,5}$ \href{https://orcid.org/0000-0002-8987-1181}{\includegraphics[scale=0.7]{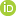}}, F. Di Cosmo$^{1,2,6}$ \href{https://orcid.org/0000-0003-0256-5913}{\includegraphics[scale=0.7]{ORCID.png}}, L. González-Bravo$^{1,7}$ \href{https://orcid.org/0000-0002-4382-7978}{\includegraphics[scale=0.5]{ORCID.png}}, \\
A. Ibort$^{1,2,8}$ \href{https://orcid.org/0000-0002-0580-5858}{\includegraphics[scale=0.7]{ORCID.png}},  G. Marmo$^{3,4,9}$ \href{https://orcid.org/0000-0003-2662-2193}{\includegraphics[scale=0.7]{ORCID.png}}}

\begin{document}

\maketitle 

\noindent
{\footnotesize $^{1}$  Department of Mathematics, University Carlos III de Madrid, Legan\'es, Madrid, Spain}  \\
{\footnotesize $^{2}$ ICMAT, Instituto de Ciencias Matem\'{a}ticas (CSIC-UAM-UC3M-UCM)} \\
{\footnotesize $^{3}$ INFN-Sezione di Napoli, Naples, Italy} \\
{\footnotesize $^{4}$ Department of Physics ``E. Pancini'', University of Naples Federico II,  Naples, Italy} \\

\bigskip
\noindent
{\footnotesize $^{5}$\texttt{fciaglia[at]math.uc3m.es} $^{6}$\texttt{fcosmo[at]math.uc3m.es}  $^{7}$\texttt{lauragon[at]math.uc3m.es} \\ $^{8}$\texttt{albertoi[at]math.uc3m.es}  $^{9}$\texttt{marmo[at]na.infn.it}}

\begin{abstract}
	An extension of Cencov's categorical description of classical inference theory to the domain of quantum systems is presented.  
	It provides a novel categorical foundation to the theory of quantum information that embraces both classical and quantum information theory in a natural way, while  also allowing to formalise the notion of quantum environment.  
	A first application of these ideas is provided by extending the notion of statistical manifold to incorporate categories, and investigating a possible, uniparametric Cramer-Rao inequality in this setting. 
\end{abstract}


\section{Introduction}

\begin{figure}[h]
	\centering
	\resizebox{9cm}{5cm}{\includegraphics{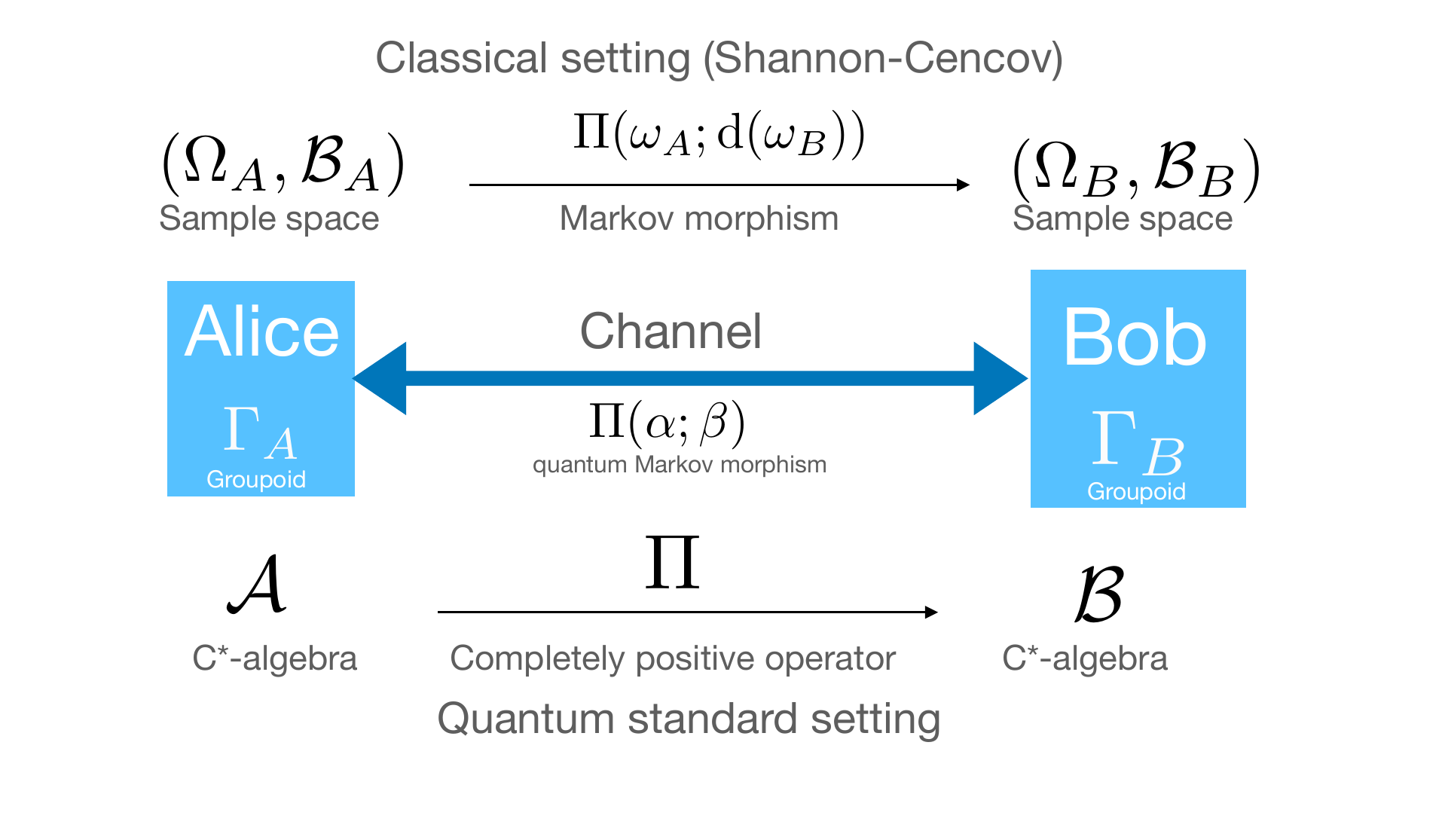}} 
	\caption{Diagrammatic representation of the fundamental problem of communication. Upper half: schematic description of classical communication among two agents; lower half: schematic description of quantum communication.}
	\label{fig:communication}
\end{figure}

This letter aims to extend Cencov's\footnote{Sometimes, the name Cencov is also spelled Chentsov.} categorical description of classical inference theory \cite{Cencov-1982} to the domain of quantum systems, providing a novel categorical foundation to the theory of quantum information.    
The main focus of information theory is to describe how information is processed and shared among various agents.
In particular, in the case of quantum information theory, agents ``live" in a quantum environment.  
The simplest schematic way of representing such exchange and manipulation of information in a purely classical environment was provided by C. Shannon in his mathematical theory of communication \cite{Shannon-1948} as illustrated in Fig. \ref{fig:communication}.
There, Shannon states: \textit{``the fundamental problem of communication is that of reproducing at one point either exactly or approximately a message selected at another point''}. 

Two agents, Alice (A) and Bob (B), share information through a physical channel (C).  
Agents A and B are mathematically modelled, in the classical Kolmogorovian setting, as certain sample spaces, say $\Omega_A$  and $\Omega_B$, carrying measurable structures given by $\sigma$-algebras of sets $\mathscr{B}_A$ and $\mathscr{B}_B$, respectively.  
Alice wants to share a message with Bob obtained by drawing random outcomes following a probability distribution law $P_A\{ d\omega \}$\footnote{Throughout this paper we will follow Cencov's notation for probability measures indicating the integral of a  random variable (measurable function) $f$ on a probability space (also called \textit{Kolmogorov space}) $(\Omega_A, \mathscr{B}_A, P_A)$ as $\int_{\Omega_A} f(\omega) P\{ d\omega \}$.}.  
The channel C is modelled as a Markov kernel (or a \textit{transition probability distribution}) $\Pi $.
A Markov kernel is a map $\Pi \colon \Omega_A \times \mathscr{B}_B \to \mathbb{R}$ such that $\Pi (\cdot, \Delta_B)$ is a measurable function on $\Omega_A$ for each measurable set $\Delta_B \in \mathscr{B}_B$, and $\Pi (\omega_A, \cdot )$ is a probability measure on $\Omega_B$ for each fixed $\omega_A \in \Omega_A$.  
We will denote the Markov kernel $\Pi$ from the measurable space $(\Omega_A,\mathscr{B}_A)$ to the measurable space $(\Omega_B,\mathscr{B}_B)$, as $\Pi \colon \Omega_A \Rightarrow \Omega_B$.

Given a Markov kernel $\Pi \colon \Omega_A \Rightarrow \Omega_B$, to any measurable function $f_B$ in $\Omega_B$, we can associate a measurable function $f_A$ in $\Omega_A$, denoted as $\Pi f_B := f_A$, as:
\begin{equation}\label{eq:class_function}
	(\Pi f_B) (\omega) = \int_{\Omega_B} \Pi (\omega, d\omega') f_B(\omega') \, .
\end{equation}
In a similar way, if $P_A$ is a probability measure on $\Omega_A$, $P_A \Pi$ is a probability measure on $\Omega_B$ defined as:
\begin{equation}\label{eq:class_states}
	(P_A \Pi) (\Delta_B) = \int_{\Omega_A} \Pi (\omega, \Delta_B) P\{ d\omega \} \, ,
\end{equation}
for any measurable set $\Delta_B$ in $\Omega_B$.   

The notation $\Pi (\omega_A, d\omega_B)$ emphasizes the stochastic nature of $\Pi$.
Indeed, in general, a Markov kernel is not induced from a measurable map $F \colon \Omega_A \to \Omega_B$  by means of the pull-back operation on functions $\Pi_F f_B = f_B\circ F$, and the push-forward operation for probability distributions $P_A \Pi_F = F_* P_A$.
When this instance presents, it is customary to  say that the Markov kernel is deterministic \cite{Fritz-2020}. 
Therefore, we can say that a general Markov kernel is inherently stochastic in the sense that it is not determined by a deterministic inference $F\colon \Omega_A \to \Omega_B$.  

To get a better understanding of the mathematical structure of the theory of information, we use Cencov's categorical conceptualisation of statistical inference theory \cite{Cencov-1982} identifying Markov kernel with classical communication channels. 
Indeed, Cencov's departing point is to assert, together with Wald, that every particular statistical problem is a problem of decision-making \cite[Preface, p. 1]{Cencov-1982}: \textit{``the statistician\footnote{We may replace `statistician' here by `agent' or `physical observer'.} having processed certain observational material, must draw conclusions as to the observed phenomenon. Since the outcome of each observation is random, one cannot usually expect there conclusions to be absolutely accurate.  It is a job for the theory to ascertain the minimal unavoidable uncertainty of the conclusions in the problem and to indicate an optimal decision rule.''}
Statistical decision rules are then  modeled as Markov kernels from a sample space $(\Omega, \mathscr{B})$ to the inference space $(\Sigma, \mathscr{S})$.    

A fundamental observation\footnote{This observation was also made by Lawvere in a seminar in 1962 \cite{Lawvere-1962}.} of Cencov's work \cite{Cencov-1965,Cencov-1978,Cencov-1982} is that the collection of all Markov kernels form a category that encodes the structural properties of statistical inference rules: \textit{``The system of all statistical decision rules, transition probability distributions, for all conceivable statistical problems, together with the natural operation of composition, forms an algebraic category. This category generates a uniform geometry of families of probabilty laws, in which the `figures' are the families and the `motions' are the decision rules''} \cite[p. vii]{Cencov-1982}. 
The composition rule for morphisms in this category is achieved by introducing a natural operation of composition of decision rules (\textit{i.e.}, transition probabilities/Markov kernels) as:
\begin{equation}\label{eq:composition}
	(\Pi_1 \circ \Pi_2) (\omega_1, A_3) = \int_{\Omega_2} \Pi_1(\omega_1, d\omega_2) \Pi_2(\omega_2, A_3) \, .
\end{equation}
The fact that this composition is indeed associative and gives rise to a category is one of the fundamental observations made by Cencov.  
Note that the units of the category are given by the deterministic Markov kernels $1_{\mathrm{id}_\Omega} \colon \Omega \Rightarrow \Omega$, associated with $\mathrm{id}_\Omega \colon \Omega \to \Omega$, the identity map.

Therefore, Cencov's category has objects measurable spaces $(\Omega, \mathscr{B})$,  morphisms given by Markov kernels (or classical channels) $\Pi \colon \Omega \Rightarrow \Omega'$, and composition law between morphisms given by Eq. \eqref{eq:composition}.    
This category is now called $\mathsf{Stoch}$, and it is the subject of a flourishing stream of research  (see, for instance \cite{Fritz-2020,F-G-P-2021,F-G-P-R-2023}, and references therein).
When the attention is shifted towards measurable spaces which are discrete and finite, we obtain the subcategory $\mathsf{FinStoch}$.
This category has been extensively used by Cencov \cite{Cencov-1982}, implicitly used by Shannon's in his mathematical model of communication, and we may argue it is also the domain where some of the most important theorems of the general theory of statistical inference have been proved.

A little twist can be applied to   $\mathsf{Stoch}$ (and $\mathsf{FinStoch}$) by considering categories whose  objects  are (finite) Kolmogorov spaces (\textit{i.e.}, finite probability spaces) $(\Omega, \mathscr{B}, P)$ instead of just (finite) measurable spaces, and whose morphisms $\Pi \colon (\Omega, \mathscr{B}, P) \Rightarrow (\Omega', \mathscr{B}', P')$ are Markov kernels $\Pi \colon \Omega \Rightarrow \Omega'$ such that $P \Pi = P'$.   
The resulting category is the classical counterpart of the category we exploit in this work, referred to as $\mathsf{NCP}$, and the motivations behind this choice are better described once the quantum case has been recalled.

In Quantum Mechanics, Shannon's original representation of the communication process takes the following form.  
Following the modern algebraic approach to quantum theories the agents Alice and Bob are quantum systems whose algebras of observables are given by unital $W^*$-algebras (\textit{i.e.}, abstract von Neumann algebras) \cite{takesaki-2002} $\mathcal{A}$ and $\mathcal{B}$, respectively.   
The states of the systems (\textit{i.e.}, the quantum analogues of classical probability measures) are normal states on the algebras, that is, linear functionals satisfying a positivity conditions (like probabilitiy measures), a normalization condition (like probability measures), and a suitable continuity condition (like probability measures dominated by a reference measure).
The transmission of information is mediated by a quantum channel described by a completely positive, unital map $\Pi$ from $\mathcal{A}$ to $\mathcal{B}$.    
Referring to standard textbooks on the subject \cite{Blackadar-2006,B-R-1987-1,takesaki-2002}, when $\mathcal{A}=\mathcal{L}^{\infty}(\Omega._{A},\mu)$ and $\mathcal{B}=\mathcal{L}^{\infty}(\Omega._{B},\nu)$, the normal states are precisely those probability measures which are absolutely continuous with respect to $\mu$ and $\nu$, respectively, and completely positive, unital maps are Markov kernels.
Therefore recover the classical case in this setting.

Enormous effort has been poured in trying to understand the basic properties governing the  situation outlined above, and, even if significant advances have been done, we believe there is still work to do to gain a complete understanding of the quantum extension of Shannon's mathematical theory of communication to the quantum setting.

Our point of view on the quantum setting is based on a recent reformulation of quantum theories in terms of groupoids and their algebras \cite{C-DC-I-M-2020,C-DC-I-M-2020-02,C-DC-I-M-S-Z-2021,C-DC-I-M-S-Z-2021-02,C-DC-I-M-S-Z-2022,C-I-M-2018,C-I-M-2019-02,C-I-M-2019-03,C-I-M-2019-05}.
This approach allows to follow closely Cencov's conceptualisation of random phenomena by describing the quantum agents, Alice and Bob, by measure groupoids $\Gamma_A$ and $\Gamma_B$ respectively (see details in Section \ref{sec:groupoids}).  
The message composed by Alice are outcomes in the space $\Omega_A$ of outputs of the system $\Gamma_A$ drawn according to the probability distribution determined by the state of the system, and the morphism that transform them into the message that B reads is determined by the natural extension of Cencov's Markov kernels, that is, a function $\Pi(\alpha_A, \alpha_B)$, whose entries are pairs of transitions $\alpha_A$, $\alpha_B$, in the groupoids $\Gamma_A$, $\Gamma_B$ respectively, such that for a fixed argument $\alpha_A$, it defines a positive type function on $\Gamma_B$,  and that for fixed argument $\alpha_B$, it defines a measurable function on $\Gamma_A$.   
The function $\Pi(\alpha_A, \alpha_B)$ transforms states into states and will be called a quantum Markov morphism (see details in Section \ref{sec:NCP2}).   
The class whose objects are couples of measure groupoids and normal states on their algebras, and morphisms are quantum Markov morphisms preserving mapping the state of the input object into the state of the output object, form an algebraic category which is the fundamental notion of our study.
In particular, we argue that the category we introduce presents a fertile environment in which to discuss Cencov's approach to classical information theory, the algebraic approach to quantum information theory, and also a possible unification of classical and quantum information geometry (a multidisciplinary field of research we now briefly recall).

Classical information geometry may be briefly (and incompletely) described as the study of the differential geometric properties of parametric models of probability distributions (\textit{e.g.}, normal distributions)  \cite{A-N-2000}.
It turns out that most of the parametric models used by applied statisticians share a nice differential geometric structure which takes the name of \textit{statistical manifold} \cite{Lauritzen-1987}.
Specifically, a statistical manifold is a smooth Riemannian manifold $(M,G)$ possessing a totally symmetric, (0,3)-covariant tensor field $T$.
The tensor $T$ allows to define a couple of affine, torsion-free connections on $M$ which are dual with respect to $G$.

In the vast majority of practical cases, the Riemannian metric  on $M$ is not ``free'', because it coincides with the so-called Fisher-Rao metric tensor $G_{FR}$ \cite{Fisher-1922,Mahalanobis-1936,Rao-1945}.
One of the biggest achievements of the marriage between probability theory, statistical theory, and information theory has been the identification of the central role played by the Fisher information and the Fisher-Rao metric tensor.
This geometrical entity provides, among many other structural insights, a lower bounds for the information content of estimators of parametric models of probability measures by means of the Cramer-Rao inequality \cite{Hendriks-1991,L-BS-W-P-D-2017}.

The Fisher-Rao metric tensor on the parameter manifold emanates from the very construction of  parametric models as subsets of probability measures opportunely parametrized by points in a manifold.  
Specifically, given a smooth manifold $M$ and a measurable space $(\Omega, \mathscr{B})$, a  model of probability measures on $(\Omega, \mathscr{B})$ parametrized by points in $M$ is nothing but an immersion map $i\colon M\rightarrow \mathcal{P}(\Omega)$, where $\mathcal{P}(\Omega)$ is the space of probability measures on $(\Omega, \mathscr{B})$, satisfying suitable regularity conditions (basically ensuring we may exploit the smooth structure on $M$ in a fruitful way) \cite{A-J-L-S-2017,A-J-L-S-2018}.
If $\{\theta^{j}\}_{j\in J}$ is a local coordinate systems on $M$, the Fisher-Rao metric tensor takes the form\footnote{Of course, the fact that equation \eqref{eqn: Fisher-Rao metric tensor} actually gives a Riemannian metric tensor on $M$ requires additional assumptions to be met by the immersion map $i$ (for instance, we are implicitly assuming that elements in the parametric models are all dominated by $\mu$).
	However, these assumptions are met in the vast majority of "applied situations".}
\begin{equation}\label{eqn: Fisher-Rao metric tensor}
	(G_{FR})_{jk}=\int_{\Omega}\frac{\partial \ln(p(x;\theta))}{\partial \theta^{j}} \frac{\partial \ln(p(x;\theta))}{\partial \theta^{k}}\,p(x;\theta)\mathrm{d}\mu(x),
\end{equation}
where $i(\theta)=p(x;\theta)\mu$.
Looking at equation \eqref{eqn: Fisher-Rao metric tensor}, we may argue that $M$ ``does not know" about $G_{FR}$ until its points are actually used to parametrize probability measures on $(\Omega,\mathscr{B})$.
Therefore, it may be conjectured that the properties of $G_{FR}$ making it an ubiquitous tool in statistical inference and estimation theory (among other fields) should be related to some inner structures of the space of probability distributions.
Indeed, we may dare to say that it is precisely this question that led Cencov to introduce the categorical framework described above.
This seemingly ``heavy'' affirmation is corroborated by the fact that one of the main results achieved in \cite{Cencov-1982} is precisely the proof that the Fisher-Rao metric tensor for parametric models on discrete and finite outcome spaces is the unique (up to a constant factor) Riemannian metric tensor satisfying an invariance property with respect to the maps in $\mathsf{FinStoch}$ having a left inverse who is also in $\mathsf{FinStoch}$ (Cencov calls these morphisms \textit{congruent embeddings}).
Moreover, Cencov also classifies all those couples of affine connections which are mutually dual with respect to $G_{FR}$ and satisfy an equivariance property with respects to congruent embeddings, thus giving a complete account of the admissible statistical manifolds for finite outcome spaces (as long as the assumption on the behaviour under congruent embeddings is imposed, of course).

Starting from the Fisher-Rao metric tensor and the notion of statistical manifold, classical information geometry takes off and leads to spectacular results in very different applied fields \cite{Amari-2016}.
Some of these results were extended also to the quantum domain, where probability distributions are replaced by quantum states \cite{BN-G-J-2003}.
In particular, one of the pillars of quantum information geometry is the celebrated Petz's theorem \cite{Petz-1996}, which states that the uniqueness of the quantum analogue of the Fisher-Rao metric tensor is necessarily lost.
This non-uniqueness, however, should be thought of as being the source of richness rather than a problem, as the wealth of different investigations and results regarding the quantum counterparts of the Fisher-Rao metric tensor seem to imply \cite{Ciaglia-2020,C-DC-L-M-M-V-V-2018,C-DN-2022,G-H-P-2009,G-I-2001,G-I-2003,G-I-2007-01,G-I-2007-02,L-R-1999}.
Moreover, the very structure of statistical manifold should be rethought in the quantum case because connections with torsions seem to be unavoidable \cite{C-DC-I-M-2022,Jencova-2001,Jencova-2003,Jencova-2003-2} (and a similar attitude seems to be fruitful also in the classical case \cite{Z-K-2019,Z-K-2019-02,Z-K-2020}).

Also along the lines of rethinking the idea of statistical manifolds, recent investigations are pointing toward a connection between information geometry and groupoids \cite{G-G-K-M-2019,G-G-K-M-2020,G-G-K-M-2023}.
This link stems from the observation that the manifold $M\times M$ is naturally a Lie groupoid whose associated Lie algebroid is $TM$, and that the appearence of the Fisher-Rao metric tensor on the parametric manifold $M$ as the ``second-order, diagonal approximation" of a relative entropy function on $M\times M$ typical of information geometry \cite{A-N-2000,C-DC-L-M-M-V-V-2018} can be naturally extended to the framework of Lie groupoid.
We find this terrain quite apt to develop a foundational investigation of the notion of statistical manifold from the perspective offered by the categorical setting we alluded to before.
In particular, by introducing the notion of statistical categories and its associated notion of statistical groupoids (see Section \ref{sec:statistical}), it is possible to look at statistical models as functors in the category introduced in this work.
In this categorical setting, an extension of Cramer-Rao inequality, that works for the classical and the quantum case simultaneously, is readily obtained.

\section{Quantum systems and groupoids}\label{sec:groupoids}

On  its most basic terms a quantum system is characterized by the outcomes $x,y,\ldots \in \Omega$ of a family of observables, and by a family of \textit{transitions}, $\alpha \colon x \to y$, experienced by the system, whose interpretation is that if the observable $A$ were measured right before the observed transition took place, the outcome would have been $x$, and if measured again, right after the transition had taken place, the result would have been $y$.   The outcome $x$ of the transition $\alpha \colon x \to y$, will be called its source, and the outcome $y$ its target. 

The natural axioms satisfied by the family of all possible transitions are those of a groupoid (see \cite{C-I-M-2019-02} for more details). In particular, transitions compose in a natural way:  the symbol $\beta \circ \alpha$ denotes the transition resulting from the occurrence first of the transition $\alpha$ and immediately afterwards, the transition $\beta$.    Two transitions $\alpha$, $\beta$ can be composed only if the target of the first coincides with the source of the second (note the backwards notation for composition), in which case they are said to be composable, and such composition law is associative.   There are unit elements, that is, transitions $1_x\colon x \to x$ such that they do not affect the transition $\alpha \colon x \to y$, when composed on the right, i.e., $\alpha \circ 1_x = \alpha$, or on the left, $1_y \circ \alpha = \alpha$, and whose physical interpretation is that the system remains unchanged during the observation and, finally, the fundamental property that implements Feynmann's principle of microscopic reversibility \cite[page 3]{Feynman-2005}, that is, for any transition $\alpha \colon x \to y$, there is another one, denoted $\alpha^{-1} \colon y \to x$, such that $\alpha^{-1} \circ \alpha = 1_x$ and $\alpha \circ \alpha^{-1} = 1_y$.   

The collection of all transitions satisfying the previously enumerated properties is called a (algebraic) groupoid $\Gamma$ with space of objects (called in what follows ``outcomes'') $\Omega$.  The map $s \colon \Gamma \to \Omega$ assigning to the transition $\alpha \colon x \to y$ the initial outcome $x = s(\alpha)$ is called the source map, and the map $t \colon \Gamma \to \Omega$ assigning to $\alpha$ its final outcome $y = t(\alpha)$ is called the target map.   We will make such structure notationally evident by writting $\Gamma \rightrightarrows \Omega$.

The previous notions provide a natural mathematical setting to Schwinger's `algebra of selective measurements' \cite{Schwinger-2000} introduced to provide an abstract setting to the foundations of atomic physics.   Transitions could also be understood in terms of the basic quantum mechanical notion of probability amplitudes because a unitary representation of the given groupoid will associate to them a family of operators directly related to the notion of probability amplitudes or `transition functions' in J. Schwinger's terminology \cite{C-I-M-2019-02}.   

It is also possible to conceive of a groupoid as an abstraction of a certain experimental setting used to describe the properties of a given system.   For instance, if we consider a charged particle moving on a certain region where detectors have been placed, the triggering of them will correspond to the possible outcomes of the system and the sequence of such triggerings would be the transitions of the system.    Another possible interpretation is offered by earlier descriptions of spectroscopic data.   Actually, as Connes suggested \cite{Connes-1994}, the Ritz-Rydberg combination principle of frequencies in spectral lines is rightly the composition law of a groupoid (in this case of a simple groupoid of pairs).   

In what follows, we will look at the groupoid used to describe a certain quantum system as a kinematical object, which means that transitions and outcomes represent just kinematical information obtained from the system without further dynamical content, that is, no specific dynamical law is associated to their description.    In this sense, we will say that the groupoid $\Gamma$ is a kimematical groupoid. It will also be said that the groupoid $\Gamma$ is the groupoid of ``configurations'' of the system, and it is associated to a specific experimental description of a quantum mechanical system (obviously, more than one kinematical groupoid can be used to describe the same quantum system, think for instance on electrons moving through bubble chambers, Stern-Gerlach devices or two-slits walls, each one of these experiments can be described by using a kinematical groupoid, all of them different).

This kinematical interpretation of the raw information provided by the experiments performed on our system must be completed with a probabilistic interpretation.  In a first step towards this aim we could extend directly Kolmogorov's mathematical description of random phenomena and we will assume that the space of outcomes $\Omega$ of the groupoid $\gamma$ is a Kolmogorov space carrying a measurable structure $\mathscr{B}$ and a probability distribution $P\{ dx \}$.   

In order to extend such structure to the whole groupoid we would be looking for a measure structure $\nu$ on $\Gamma$ such that  it is consistent with the projection maps $s,t$ and the action of the groupoid $\Gamma$ on itself by left or right translations.   It turns out that such structure was analized separately by A. Connes \cite{Connes-1979} and R. Haar \cite{Hahn-1978,Hahn-1978-2} following the lead by G.W. Mackey and others.  

As it turns out, the measure theoretical structure of the groupoid $\Gamma \rightrightarrows \Omega$ is largely determined by the probability measure $P$ on $\Omega$, a left-invariant family of Haar measures, that is, a family of measures $\nu^x$ with support in $\Gamma^x = t^{-1 (x)}$, satisfying that $\alpha \nu^x = \nu^y$ for any $\alpha \colon x \to y$ (also called a transverse function in Connes' terminology), and $(\alpha \nu^x) (\Delta) = \nu^x (\alpha^{-1}\circ \Delta)$, and, finally a modular map $\delta\colon \Gamma \to \mathbb{R}^+$ which is a homomorphism of groupoids.  These ingredients determine a essentially unique measure $\nu$ in $\Gamma$ that disintegrates as $\nu = \int_\Omega \nu^x P\{ dx \}$ and such that the Radon-Nikodym derivative of the measure $\nu^{-1}$ defined as $\nu^{-1}(\Delta) = \nu (\Delta^{-1})$, with respect to $\nu$ is the integrable modular map $\delta$.  
Under minimal requirements on the measure theoretical properties of the spaces, for instance that they are standard Borel spaces, the groupoid $\Gamma \rightrightarrows \Omega$ together with the class of the measure $\nu$ has associated a von Neumann algebra $\nu (\Gamma) $ with support Hilbert space $L^2(\Gamma, \nu)$ \cite{Hahn-1978-2}.    The pair $(\Gamma\rightrightarrows \Omega, [\nu])$ will be called a measure groupoid and it constitutes a \emph{bona fide} extension of Kolmogorov's model to describe random phenomena when we consider boht, random outcomes and transitions among them.   

A relevant observation here is that the measure $\nu$ does not have to be interpreted in statistical terms, that is, it does not provide the statistical frequencies of events in $\Gamma$.    Such statistical interpretation could be carried by a more sophisticated notion of measure, e.g. a grade 2-measure as argued in \cite{Sorkin-1994} (see, for instance, \cite{C-I-M-2019-05,C-DC-2023} for a detailed discussion) or, as it is commonly done, by identifying the physical states of the theory with the states of the von Neumann algebra $\nu (\Gamma)$ of the groupoid.   This is the point of view that will be taken here.

A large class of topological groupoids $\Gamma$, for instance those that are locally-compact, Hausdorff topological spaces for which $s,t$ , as well as $x\mapsto 1_{x}$, $\alpha\mapsto\alpha^{-1}$, and $(\beta,\alpha)\mapsto\beta\circ\alpha$, are continuous maps has been already considered in the literature. When $\Gamma$ and $\Omega$ are smooth manifolds and all the maps are smooth (in particular, $s,t$ are smooth submersions), we say that the groupoid is a {\itshape Lie groupoid}.   In what follows, and in order to avoid technical complications in the exposition, we will assume that the groupoids used in the description of quantum systems are just countable discrete.  In such case the Borel structures are just given by the family of all subsets, all measures have atoms given by the elements of the groupoid itself.  Then the von Neumann algebra of the measure groupoid $(\Gamma, \nu)$ is just the completion in the weak (or strong) topology of the left regular representation of the abstract algebra $\mathbb{C}[\Gamma]$ of the groupoid.  Elements $a \in \mathbb{C}[\Gamma]$ are finite formal linear combinations of transitions $\alpha\in \Gamma$, $a = \sum_\alpha a_\alpha \alpha$, all $a_\alpha \in \mathbb{C}$ are zero except for a finite number of them, and $A = \lambda (a)$ is the bounded operator on $L^2(\Gamma, \nu)$ given by :
$$
(A\Psi) (\beta) = (\lambda (a) \Psi) (\beta) =  \sum_{t(\alpha) = t(\beta)} a_\alpha \delta^{1/2}(\alpha) \Psi(\alpha^{-1}\circ \beta) \, .
$$
It is a trivial exercise to check that the assignment $a \in \mathbb{C}[\Gamma] \mapsto A = \lambda(a) \in \mathscr{B}(L^2(\Gamma, \nu))$ is a $*$-algebra representation of the $*$-algebra $\mathbb{C}[\Gamma]$ in the $C^*$-algebra of bounded operators on $L^2(\Gamma, \nu)$, and then:
$$
\nu (\Gamma) = \overline{\lambda(\mathbb{C}[\Gamma])}^{\mathrm{WOP}} \, .
$$

As indicated before physical states of the system described by the groupoid $\Gamma$ will be identified with states $\rho$ on the von Neumann algebra of the groupoid, that is normalised positive functionals $\rho \colon \nu (\Gamma) \to \mathbb{C}$.  Then, any state has a characteristic function $\varphi \colon \Gamma \to \mathbb{C}$ associated to it by restriction of $\rho$ to $\Gamma$, that is:
$$
\varphi (\alpha) = \rho (\lambda (\alpha)) \, .
$$
In such situation if $A\in \nu(\Gamma)$, there is a sequence $a_n\in \mathbb{C}[\Gamma]$ such that $\lambda(a_n) \to A$, and:
\begin{equation}\label{eq:characteristic}
	\rho (A) = \lim_n \rho(\lambda(a_n)) = \lim_n \sum_\alpha a_n(\alpha) \varphi (\alpha) \nu (\alpha)\, .
\end{equation}
Characteristic functions $\varphi$ associated to states are positive definite, that is they satisfy the positivity property: for all $N \in \mathbb{N}$, $\zeta_k\in \mathbb{C}$, $k = 1, \ldots, N$, $\alpha_k \in \Gamma$, then 
$$
\sum_{t(\alpha_k) = t(\alpha_l)} \bar{\zeta_k} \zeta_l  \varphi (\alpha_k^{-1} \circ \alpha_l ) \geq 0 \, .
$$
Moreover, if $\varphi$ is the characteristic function of the state $\rho$, then we have the normalisation condition: $\sum_{x\in \Omega} \varphi(1_x) P(\{x\}) = 1$, resulting form $\rho (\mathbf{1}) = 1$.   Clearly the numbers $\varphi (1_x)$ are non-negative real numbers, hence they determine a probability distribution $p(x) = \varphi (1_x) P(\{ x \})$ in the space of outcomes $\Omega$.  
Conversely, because Eq. (\ref{eq:characteristic}) any normalised positive definite function $\varphi$ on $\Gamma$ will define a state on the von Neumann algebra of $\Gamma$.  

Of course if the measure groupoid $\Gamma$ is just the trivial groupoid defined by the set $\Omega$ carrying a Kolmogorov structure, then its von Neumann algebra is just the Abelian von Neumann algebra of essentially bounded functions on $\Omega$, $\nu (\Gamma) = L^\infty(\Omega, P)$, and the states of the theory are just probability measures $p$ on $\Omega$ absolutely continuous with respect to $P$.   It is remarkable that any Abelian von Neumann algebra has this form, hence corresponding to trivial measure groupoids.   

The simplest non-trivial situation corresponds to $\Gamma$ being the groupoid of pairs of a countably discrete space $\Omega$.  If $\Omega$ is finite, then the von Neumann algebra $\nu (\Omega)$ can be identified with the algebra of $N\times N$ matrices $M_N(\mathbb{C})$, with $N = |\Omega|$.    If $\Omega$ is infinite, then again the von Neumann algebra of the goupoid of pairs $\Gamma = \Omega \times \Omega \rightrightarrows \Omega$, can be identified with the factor of Type $I_\infty$ of all bounded linear operators on $L^2(\Omega,P)$,   and the states of the theory will be described by density operators because of Gleason's theorem.   

Much more complicated situations emerge readily by considering the composition of families of finite systems (see, for instance \cite{C-DC-F-I-K-M-2023} where the groupoidal analysis of Powers construction of Type III factors was discussed) or by considering countably infinite systems with a finite space of outcomes in which case we will be describing quantum systems by groupoids whose von Neumann algebra would be Type II factors.


\section{The NCP category and quantum environments}\label{sec:NCP2}

As anticipated in the introduction, one of the main purpose of this work is to introduce the category $\mathsf{NCP}$ that allows to deal with classical and quantum information theory in a way that incorporates both Cencov's ideas about the role of $\mathsf{Stoch}$ and $\mathsf{FinStoch}$ in classical statistics,  the  algebraic approach to quantum information theory, and the groupoidal point of view discussed above.
We start considering countably discrete groupoids $\Gamma\rightrightarrows \Omega$ and their groupoid von Neumann algebras $\mathcal{M}$ as in the end of the previous section.
Then, because of the identification between states and normalized, positive definite functions on $\Gamma$, we can introduce the notion of quantum Markov kernel in analogy with the classical case as a map: $\Pi \colon \Gamma_1 \times \Gamma_2 \to \mathbb{C}$, such that:

\begin{enumerate}
	\item[i.-] Normalization: $\sum_{x\in \Omega_2}\Pi (\alpha_1, \mathbf{1} )= 1$.
	\item [ii.-] Positivity: $\Pi (\alpha_1, \cdot )$ is a positive definite function on $\Gamma_2$, for every $\alpha_1 \in \Gamma_1$.
	\item[iii.-] Hermiticity: $\overline{\Pi(\alpha_1,\alpha_2)} = \delta(\alpha_2) \Pi (\alpha_1^{-1}, \alpha_2^{-1})$.
\end{enumerate}
Defining 
\begin{equation}\label{eqn: quantum markov kernels on states}
	(\varphi_1\Pi) (\alpha_2) = \int_{\Gamma_1} \varphi_1 (\alpha_1) \Pi (\alpha_1, \alpha_2) \nu_1 \{ d\alpha_1\} \, ,
\end{equation}
in close analogy with  \eqref{eq:class_states}, and 
\begin{equation}\label{eqn: quantum markov kernels on observables}
	\Pi f_2(\alpha_1) = \int_{\Gamma_2} \Pi (\alpha_1, \alpha_2) \nu_2 \{ d\alpha_2\} \, ,
\end{equation}
similarly to \eqref{eq:class_function}, it is a matter of direct computation to show that $\varphi_2$ is a positive definite function on $\Gamma_{2}$ if $\varphi_1$ is so, and that $\overline{(\Pi f_2)} = \Pi f_2$, provided that $\overline{f_2} = f_2$.
A composition law can be defined according to
\begin{equation}\label{eqn: composition law for groupoid quantum markov kernels}
	(\Pi_{12}\circ \Pi_{23})(\alpha_{1},\alpha_{3}):=\int_{\Gamma_{2}}\Pi_{12}(\alpha_{1},\alpha_{2})\,\Pi(\alpha_{2},\alpha_{3})\, \nu_2 \{ d\alpha_2\}.
\end{equation}
The associativity of this composition rule depends on the $\sigma$-additivity of the measure $\nu$, which must be assumed in order to build the groupoid von Neumann algebra $\nu(\Gamma)$ in the first place \cite{Connes-1979,Kastler-1982}.
Then, a category can be built using couples $(\Gamma\rightrightarrows \Omega, \varphi)$, where $ \Gamma\rightrightarrows \Omega$ is a countable discrete groupoid, and $\varphi\colon \Gamma \to \mathbb{C}$ is a normalized positive definite function on $\Gamma$, as objects, and quantum Markov kernels, denoted by $\Pi \colon (\Gamma_1, \varphi_1) \Rightarrow (\Gamma_2, \varphi_2)$, and satisfying the additional property  $(\varphi_1\Pi)=\varphi_{2}$, as morphisms.
The category thus built is reminiscent of $\mathsf{Stoch}$ and $\mathsf{FinStoch}$, and is the starting point for the definition of the category $\mathsf{NCP}$ alluded to in the introduction.

In order to explicitly define $\mathsf{NCP}$, let us start noting that the couple $(\Gamma\rightrightarrows \Omega, \varphi)$ can be algebraically described through the groupoid von Neumann algebra $\nu(\Gamma)$ and the state on it determined by the positive definite function $\varphi$.
Moreover, we recall that a quantum Markov kernel $\Pi$ gives rise to two additional maps by means of equation \eqref{eqn: quantum markov kernels on states} and equation \eqref{eqn: quantum markov kernels on observables}.
From the algebraic point of view, equation \eqref{eqn: quantum markov kernels on states} gives rise to a linear map that sends states on the groupoid von Neumann algebra $\nu(\Gamma_{1})$ to states in the groupoid von Neumann algebra $\nu(\Gamma_{2})$.
On the other hand, equation \eqref{eqn: quantum markov kernels on observables} gives rise to a linear map between $\nu(\Gamma_{2})$ and $\nu(\Gamma_{1})$ which preserves self-adjoint elements, and also positive ones.
Also, note how these two linear maps ``flow in opposite directions''.

Putting everything together, we define the category $\mathsf{NCP}$ as that category whose objects are couples $(\mathcal{M},\rho)$, with $\mathcal{M}$ a $W^{*}$-algebra and $\rho$ a normal state on it, and  whose morphisms $\Pi\colon(\mathcal{M},\rho)\Rightarrow (\mathcal{N},\sigma)$ are couples $(f,f_{*})$, where $f\colon \mathcal{N}\rightarrow\mathcal{M}$ is a normal, completely positive, unital map, and $f_{*}\colon \mathcal{M}_{*}\rightarrow\mathcal{N}_{*}$ is the predual map of $f$ satisfying the additional compatibility condition $f_{*}(\rho)=\sigma$.
The proof that $\mathsf{NCP}$ is indeed a category is a matter of direct inspection, and its basically due to the associativity of $f$ and $f_{*}$ (see \cite{C-DC-GB-2023} for more technical details on this category).

A few remarks are in order here.
First of all, the choice of working with $W^{*}$-algebras (\textit{i.e.}, $C^{*}$-algebras which are the Banach dual  of a Banach space, the so-called predual) with separable predual, and normal states on them is driven the fact that this is the situation encountered  in the vast majority of ``applications'', and by the nice mathematical properties possessed by these objects \cite{Blackadar-2006,B-R-1987-1,takesaki-2002}.
Note  that using $W^{*}$-algebras allows to deal with  classical and quantum information geometry simultaneously, as already remarked in \cite{C-J-S-2020,C-J-S-2020-02,C-J-S-2022,C-DN-J-S-2023}.
Indeed, Abelian $W^{*}$-algebras like $\mathcal{L}^{\infty}(\Omega,\mu)$ are a perfect environment to discuss classical situations, as commented in the introduction.
Moreover, let us remark that the notion of morphism introduced in $\mathsf{NCP}$ captures also classical Markov kernels.

The normality assumption on $f$ amounts to the existence of its predual map $f_{*}$, while the requirement of complete-positivity \cite{Choi-1975,Stinespring-1955} is essentially driven by the need of considering tensor products when dealing with composite systems.
In the remaining of this work, this enhanced positivity condition will not play a relevant role, even though the characterisation of quantum channels satisfying complete positivity is a major problem.

\vspace{0.5cm}

In specific situations, we would like to identify a given number of agents and channels processing information instead of the full huge category $\mathsf{NCP}$.  
This is readily done by introducing the notion of a \textbf{quantum environment}, that is, a family of agents working in their laboratories with their corresponding physical systems and communication channels among them. 
Note that we do not consider separately classical communication channels and quantum ones, in the same way that we do not treat separately classical systems and proper quantum ones, because the notions introduced before allow us to consider all together once for all.
In other words a ``quantum environment'' can be loosely defined as a family of quantum and classical systems together including the interactions and the processes taken place among them.   

The notions we have introduced before allow for a natural formalisation of this important concept.
Because of the analysis carried on in the previous section we now understand that both classical and quantum systems can be properly described by using groupoids and their algebras, where classical ones will have associated Abelian von Neumann algebras, and proper quantum ones will have associated the von Neumann algebras of their corresponding groupoids. 
The processes taking place among them will be described by  morphisms in $\mathsf{NCP}$, and they could include both classical Markov kernels among classical systems, proper quantum channels among quantum ones, or mixed situations.  
In all these cases,  processes will describe exchange and manipulation of information among the various agents present in the given environment.   
Finally, it is clear that if two processes are present in the environment, and they are composable, their composition should also be a possible process taking place in the environment.

Therefore, we may conclude that a quantum environment is closed under composition of morphisms in $\mathsf{NCP}$, or, more precisely, a quantum environment $\mathbf{Q}$ will be defined as a small subcategory of $\mathsf{NCP}$ such that the $W^{*}$-algebra  appearing in it will be groupoid von Neumann algebras.

A few observations are in order here.  Both the condition that the category $\mathbf{Q}$ defining a quantum environment is small and that the von Neumann algebras must be groupoid algebras could be dispensed with from a purely formal mathematical perspective.  
Indeed, there is no reason, apart from an assumption running in the background of our thoughts about the intelligibility of our universe, that makes us to believe that in order to describe natural phenomena we can restrict our mathematics to that provided by set theory.   
It might very well happen that Nature is inherently non set-theoretical and our use of set theoretical notions is a prejudice derived from our own historical development.    
On the other side, there are not sufficient experimental evidence that would support such radical departing from the standard use of mathematics in Physics.  
So, together with E. P. Wigner \cite{Wigner-1960}, we will wonder about the unreasonable effectiveness of (set-theoretical) mathematics and will keep the categories used in our description of physical systems small.   

Concerning the second assumption, let us remark that every $W^{*}$-algebra $\mathcal{M}$ has associated a groupoid $\mathscr{G}(\mathcal{M})$, the groupoid whose objects are projections $p \in \mathcal{M}$, and morphisms partial isometries among them.    
Under suitable technical conditions,  the algebra $\mathcal{M}$ can be thought as a quotient of the $C^*$-algebra of $\mathscr{G}(\mathcal{M})$.  
In particular, when the algebra $\mathcal{M}$ is finite-dimensional, there will always be a groupoid of which $\mathcal{M}$ is the associated groupoid von Neumann algebra \cite{I-R-2019}.
However, this mathematical argument will not provide a natural, direct, interpretation of the elements of such groupoid in physical terms as argued in Section \ref{sec:groupoids}.   
In any case, looking for a relation as close as possible between mathematical structures and the assignment of physical meaning we are expected to provided for them, it would be under reason to stick to the assumptions established before even if for a large part of the mathematical arguments considered they are not strictly necessary.

\section{Statistical categories and the Cramer-Rao inequality}\label{sec:statistical}

As it was argued in the introduction, parametric models allow to introduce new analytical and geometrical tools to study the system of interest.
Most importantly the Fisher-Rao metric and its derived geometrical notions.   
At its most basic level, we can say that a parametric model is just a smooth manifold $\Sigma$ and an injective map $i \colon \Sigma \to \mathcal{P}(\Omega)$, with $\mathcal{P}(\Omega)$ denoting the family of all probability distributions on a measurable space $(\Omega,\mathscr{B})$.  
The map is also required to be smooth with respect to the smooth structure of the Banach space of signed measures on $(\Omega,\mathscr{B})$ with bounded total variation in which $\mathcal{P}(\Omega)$ naturally sits \cite{A-J-L-S-2017,A-J-L-S-2018}.
The probability distribution $i(\theta) \in \mathcal{P}(\Omega)$ is denoted typically as $p(\theta)$ or $p_\theta$, for any $\theta \in \Sigma$.  

From our previous discussion, this parametric description of random phenomena lacks a fundamental ingredient, the possible Markov kernels relating two probabilities $p = p(\theta)$ and $p' = p(\theta')$ with each other.     
Specifically, parametric models consider a parametrization for the probability distributions in terms of $\Sigma$, but  do not consider the analogue notion for the morphisms of interest, that is for the family of channels, either classical or quantum, that are considered to be relevant in the system under study.  
How to encode this additional information is precisely what we address in what remains of this discussion.   

It is just natural to consider that, because the algebraic structure of a quantum environment is that of an algebraic category, an adequate parametric model for a family of states and channels would be a category $\mathbf{C} \rightrightarrows \Sigma$ whose objects $\theta \in \Sigma$ will be modelling the states $\rho$ of the system, and whose morphisms $\alpha \colon \theta \to \theta'$ would be modelling the channels $\Pi$ of the quantum environment.   
Moreover, this assignment should respect the algebraic properties of the components, that is, if $\alpha\colon \theta \to \theta'$ models a channel $\Pi$ and $\alpha' \colon \theta' \to \theta''$, then, $\alpha' \circ \alpha \colon \theta \to \theta''$ should model the composition of the two channels $\Pi \circ \Pi'$, and, in addition, it should assign units $1_\theta$, to trivial channels.   
In other words, the assignment sending objects and morphisms from the model category $\mathbf{C} \rightrightarrows \Sigma$ into objects  and  morphisms in the category $\mathbf{Q}$ (or even $\mathsf{NCP}$) must be a functor among categories.

Moreover, in order to perform a geometrical analysis of the properties of the model, it is natural to assure that the category $\mathbf{C}$ is a smooth manifold.   
More precisely, we will assume that the category $\mathbf{C} \rightrightarrows \Sigma$ is a Lie category.   
A Lie category is a small category  $\mathbf{C} \rightrightarrows \Sigma$ such that $\mathbf{C}$ is a smooth manifold (possibly with boundary), $\Sigma$ is a smooth manifold without boundary, the source and target maps $s,t \colon \mathbf{C} \to \Sigma$, are smooth submersions and both, the composition map $m \colon \mathbf{C}^{(2)} \to \mathbf{C}$, where $\mathbf{C}^{(2)}$ is the set of pairs of composable morphisms and $m(\alpha , \beta) = \alpha \circ \beta$, and the map $i \colon \Sigma \to \mathbf{C}$, $i(x) = 1_x$, are smooth maps.  
Moreover, if $\mathbf{C}$ has a non-empty boundary, we will assume that the restrictions of the source and target maps to it, are again smooth submersions.   
Given $\theta\in \Sigma$, we will denote by $\mathbf{C}^\theta = \{ \beta \colon \varphi \to \theta \}$, and, similarly, $\mathbf{C}_\theta = \{ \beta \colon\theta \to \varphi  \}$.
We refer to \cite{Grad-2023} for a recent account on more technical aspects of Lie categories.

Consider, for instance, the partial order category associated to the standard partial order in $\mathbb{R}$, that is $\mathbf{C} = \{ (x,y) \mid x \leq y \}$, with the composition law induced from the groupoid of paris of $\mathbb{R}$, that is, $(x,y) \circ (y,z ) = (x,z)$.   
Then, $\mathbf{C} \rightrightarrows \mathbb{R}$ is a Lie category with source and target maps $s(x,y) = y$, $t(x,y) = x$.   
Moreover $\partial \mathbf{C} = \{ (x,x) \mid x \in \mathbb{R} \}$, is non-empty and the restrictions of $s,t$ to it are smooth submersions.

There is a natural notion of ``infinitesimal elements'' in Lie categories.
We consider the Lie algebra of invariant vector fields $X$ on the manifold $\mathbf{C}$ with respect to the  left (or right) action of the category $\mathbf{C}$ on itself.
Specifically, for any $\alpha\colon \theta \to \theta'$, consider the smooth map $L_\alpha \colon \mathbf{C}^{\theta} \to \mathbf{C}^{\theta'}$, given by $L_\alpha (\beta) = \alpha \circ \beta$, then the vector field $X$ is left-invariant if for any $\alpha \colon \theta \to \theta'$, then $TL_\alpha X(\beta) = X (\alpha \circ \beta)$, for any $\beta\in \mathbf{C}^\theta$.   
The restriction of left-invariant vector fields $X$ to the submanifold $\Sigma$ defines a vector bundle that will be called the (left) Lie algebroid of $\mathbf{C} \rightrightarrows \Sigma$ and denoted as $A_L(\mathbf{C})$.  
In a similar way, we can define the right Lie algebroid of $\mathbf{C}$ by using right-invariant vector field.  
It can be shown that, if the units of the category $\mathbf{C}$ are interior, than both the left and the right Lie algebroids of the category are isomorphic and they agree with the Lie algebroid of the Lie groupoid of the category \cite{Grad-2023}.  

In what follows, we will refer always to the left Lie algebroid of the category $\mathbf{C}$.  
The space of smooth sections $\xi$ of the Lie algebroid $A(\mathbf{C})$ carries a canonical Lie bracket induced from the Lie bracket of the corresponding invariant vector fields and there is a natural exponential map 
$\mathrm{Exp\,} \colon \Gamma (A(\mathbf{C})) \times \Sigma \to \mathbf{C}$, assigning to any such section $\xi$ a family of submanifolds in $\mathbf{C}$ defined as follows:
\begin{equation}\label{eq:exp}
	\mathrm{Exp} ( s\xi, \theta) = \varphi_s^{X_\xi}(\theta) \, ,
\end{equation}
for $\theta \in \Sigma$, $s \in (-\epsilon, \epsilon)$, for some $\epsilon > 0$, and $\varphi_s^{X_\xi}$ denotes the flow of the left-invariant vector field $X_\xi$ associated to $\xi$.

We will define a statistical category as a Lie category $\mathbf{C} \rightrightarrows \Sigma$ together with an injective functor $i \colon \mathbf{C} \to \mathsf{NCP}$, that is, a map that assigns to any element $\theta \in \Sigma$, an object $i(\theta) = (\mathcal{M}_\theta, \rho_\theta)$ in $\mathsf{NCP}$, and to any morphism $\alpha \colon \theta \to \theta'$, a   morphism $\Pi (\alpha) \colon (\mathcal{M}_\theta, \rho_\theta)\Rightarrow (\mathcal{M}_{\theta'}, \rho_{\theta'})$, such that $\Pi (\alpha \circ \beta) = \Pi (\beta) \circ \Pi(\alpha)$, and $\Pi (1_\theta) = 1_{\rho_\theta}$.     
Therefore , the interpretation of the statistical category for a quantum environmnet $(\mathbf{C}, i, \mathbf{Q})$ is that it provides a smooth parametric model both for a family of quantum states relevant for the problem at hand together with a family of quantum channels among them.  
Note that a quantum environment $\mathbf{Q}$ which is itself a Lie category can be considered a statistical category.     


Note that any statistical category extends the notion of a statistical manifold typical of information geometry.  Indeed, smooth manifolds $\Sigma$ can be considered to be Lie categories, albeit quite ``dumb'' ones because the only morphisms are the units $1_\theta \colon \theta \to \theta$, $\theta \in \Sigma$.  
Hence, a statistical manifold $(\Sigma, i, \mathcal{P}(\Omega))$ is a statistical category whith quantum environment provided by the family of Abelian $W^{*}$-algebras $L^\infty(\Omega, P)$, with $(\Omega, \mathscr{B},P)$ a Kolmogorov space.   
Moreover, if $(\mathbf{C}, i, \mathbf{Q})$ is a statistical category in which $\mathbf{C}$ is as before just a manifold,  we may consider the assignment $\theta \mapsto \rho_\theta$ as a smooth model for the family of states $\rho \in \mathscr{S}(\mathbf{Q})$ as discussed, for instance in \cite{C-DC-I-M-2017}.  


Any Lie category $\mathbf{C} \rightrightarrows \Sigma$ contains a Lie groupoid $G \rightrightarrows \Sigma$, consisting of all its invertible morphisms.   
In this sense, any statistical category provides a statistical groupoid by restriction of the functor $\Pi$.   
Indeed, the functorial properties of the assignment $\alpha \mapsto \Pi (\alpha)$, imply that $\Pi (\alpha^{-1}) = \Pi (\alpha)^{-1}$ and the morphism $\Pi (\alpha)$ corresponding to an invertible morphism $\alpha \in G$, is invertible in the category $\mathbf{Q}$, that is, there is another morphism $\Pi' \colon (\rho' , \mathcal{M}') \to (\rho, \mathcal{M})$ such that $ \Pi'\circ \Pi  $ is the unit morphism at $(\rho, \mathcal{M})$ and $ \Pi \circ \Pi' $ is the unit morphism at $(\rho' , \mathcal{M}')$. 
Lie groupoids in the context of the geometry of information theory were introduced by K. Grabowska, J. Grabowski, M. Kus and G. Marmo \cite{G-G-K-M-2019,G-G-K-M-2020,G-G-K-M-2023}, and remarkable geometric information concerning the structure of divergence functions and other geometrical structures were derived.   
We believe the notions and ideas presented in this note provide additional support for the relevance of groupoids and categories in the context of information theory, in general, and information geometry, in particular. 
These and other related aspects will be presented elsewhere.

\vspace{0.5cm}

We now turn our attention toward the use of statistical categories to obtain a version of the Cramer-Rao bound, for a single estimator, which is adapted to our setting, and is essentially connected with the Gelfand-Naimark-Segal (GNS) representation associated with a given state.
For this purpose we will follow and adapt the derivation of the standard uniparametric quantum Cramer-Rao inequality \cite{Helstrom-1967,Helstrom-1976,Petz-2002,Y-L-1973}.  

Let $\theta_0 \in \Sigma$ be a fixed point in our space of parameters of the model, and $\rho_0 = \rho (\theta_0)$ be the corresponding state in the $W^{*}$-algebra $\mathcal{M}_0$.    
We now briefly recall the GNS representation determined by the state $\rho_0$ \cite{B-R-1987-1,C-M-2009}.
Consider the so-called \textit{Gelfand ideal} generated by $\rho_{0}$, that is, the left ideal $\mathcal{J}_0 = \{ A \in \mathcal{M}_0 \mid \rho_0(A^*A) = 0 \}$.
Define the GNS Hilbert space $\mathcal{H}_0$ associated with $\rho_{0}$ as the Hilbert space obtained by completing the quotient space $\mathcal{M}_0/ \mathcal{J}_{0}$,  with respect to the norm associated to the inner product $\langle A \mid B \rangle_0 = \rho (A^*B)$, where $|A \rangle = A + \mathcal{J}_0$, denotes the vector associated to $A\in\mathcal{M}_0$ in $\mathcal{A}/\mathcal{J}_0$.
Then, the GNS representation $\pi_0 \colon \mathcal{M}_0 \to \mathscr{B}(\mathcal{H}_0)$ is the homomorphism of $C^*$-algebras defined as $\pi_0(A) |B\rangle = |AB\rangle$, where $|B \rangle \in \mathcal{H}_0$.

Consider now the folium $\mathcal{W}_0$ of the state $\rho_0$, that is, all those states $\rho$ on $\mathcal{M}_{0}$ such that there is a density operator\footnote{A density operator on a Hilbert space $\mathcal{H}$ is a trace-class, positive semidefinite operator with unit trace.} $D$ on $\mathcal{H}_0$ satisfying $\rho (A) = \mathrm{Tr\,}(D \pi_0 (A))$.  
Suppose that the statistical category $\mathbf{C}$ satisfies the condition $i(\Sigma) \subset \mathcal{W}_0$, which means that the states we are modelling using the statistical category $\mathbf{C}$ lie in the folium of $\rho_0$.
In particular, this means there exists a family of density operators $D(\theta)$ such that $\rho_\theta (A) = \mathrm{Tr\,} (D(\theta) \pi_0 (A))$.

Given an ``infinitesimal element'' of the statistical category $\mathbf{C}$, that is, a cross section $\xi$ of its Lie algebroid $\pi \colon A(\mathbf{C}) \to \Sigma$, there is a one-dimensional family of states $\rho_{s}:=i(\exp (s \xi_0))$, where $\exp (s \xi_0)$ denotes the projection on $\Sigma$ of the curve defined by the exponential map on the Lie algebroid $A(\mathbf{C})$ (see Eq.  \eqref{eq:exp}), namely, $\exp (s \xi_0) : = t ( \mathrm{Exp\,} (s \xi (\theta_0))$.
An element $A \in \mathcal{M}_0$ is an unbiased estimator for $s$, if the expected value of $A$ on $\rho_{s}$ is essentially $s$:
\begin{equation}\label{eqn: unbiased estimator}
	\rho_{s} (A) = s \, , \qquad s \in (-\epsilon, \epsilon) \, , \quad \epsilon > 0 \, .
\end{equation}
The idea of interpretating $A$ as an estimator follows from the fact that equation \eqref{eqn: unbiased estimator} implies that experimental observations can be used to infer  the value of the parameter $s$ itself.

The map $\Phi_\xi \colon \mathcal{H}_0 \to \mathbb{C}$,  defined by 
$$
\Phi_\xi |B\rangle = \left.\frac{\partial}{\partial s}\right|_{s = 0} \rho_{\exp (s \xi_0)} (B) = \mathrm{Tr\,} (D(\exp (s \xi_0)) \pi_0( B)) \, ,
$$
is a continuous linear map on $\mathcal{H}_0$. 
Therefore, Riesz's theorem implies there is a unique element  $\ell_\xi \in \mathcal{H}_0$ such that
$$
\Phi_\xi (B) = \langle \ell_\xi \mid B \rangle_0 \, .
$$
In particular, if $A$ is an unbiased estimator, it follows from equation \eqref{eqn: unbiased estimator} that
$$
\langle \ell_\xi \mid A \rangle_0=\Phi_\xi (A)=\left. \frac{\partial}{\partial s}\right|_{s = 0} \rho_{\exp (s \xi_0)} (A)  = 1 \, .
$$
Consequently, we have
$$
1 = | \langle \ell_\xi \mid A \rangle_0 | \leq || \ell_\xi ||_0 || A ||_0 \, ,
$$
which is equivalent to
\begin{equation}\label{eqn: cramer-rao bound}
	\rho_0 (A^*A) = \langle A \mid A \rangle_0 \geq \frac{1}{\langle \ell_\xi \mid \ell_\xi \rangle_0} \, .
\end{equation}
Equation \eqref{eqn: cramer-rao bound} consitutes the generalised Cramer-Rao inequality we were looking for.    
Following Petz, the term $\rho_0 (A^*A)$ is interpreted as a generalized statistical  variance  of $A$ on the state $\rho_{0}$, and equation \eqref{eqn: cramer-rao bound} shows it is bounded below by the inverse of $\langle \ell_\xi \mid \ell_\xi \rangle_0$, a quantity that does not depend on $A$, but only on the infinitesimal object $\xi$.
Note that  $\langle \ell_\xi \mid \ell_\xi \rangle_0$ may be interpreted as a pointwise inner product on the Lie algebroid $A(\mathbf{C}) \to \Sigma$, so that, when suitable additional regularity conditions are met, we can define a sort of Fisher-Rao metric on the statistical category $\mathbf{C}$ setting
$$
G_F(\xi, \zeta) = \langle \ell_\xi \mid \ell_\zeta \rangle_0 \, .
$$
Because of the appearence of the GNS Hilbert product in the definition of $G_{F}$, and motivated by the discussion in section 6 of \cite{C-J-S-2020}, we conjecture this metric to be the analogue, in the context of statistical categories, of the Bures-Helstrom metric tensor \cite{Helstrom-1967,Helstrom-1976} (also discussed, from a perspective different from that of estimation theory, in  \cite{B-C-1994,Cantoni-1975,Cantoni-1977,Dittmann-1995,Uhlmann-1976,Uhlmann-2011}).

Of course, this brief discussion of the Cramer-Rao inequality for one-dimensional estimators in the setting of statistical categories is only a preliminary step toward a  systematic development of (multiparametric) estimation theory for statistical categories.
Indeed, the very fact of dealing with possibly non-commutative algebras immediately leads to the appearence of a zoo of possible quantum counterparts of covariances \cite{G-H-P-2009,Petz-2002},  and of the Fisher-Rao metric tensor \cite{C-M-1991,L-R-1999,Petz-1996}, and a careful comparison of all these possibilities must be provided.
We believe the covariance associated to the GNS Hilbert product will still provide the best version of the Cramer-Rao inequality (very much like it happens in quantum information theory for finite-level quantum systems described by type $I_{n}$  factors \cite{Fuchs-1996}), but the subleties of multiparameter quantum estimation theory \cite{L-Y-L-W-2020,Paris-2009,Suzuki-2019,S-Y-H-2020} call for a more detailed discussion we aim to present elsewhere.


\section{Conclusions and discussion}

A new categorical background to analyse quantum information theory has been presented.   It extends in a natural way Cencov's categorical presentation of statistical inference theory and the standard description of quantum information theory in terms of algebras of operators and quantum channels.   The lousy notion of quantum environment  can be formulated precisely in terms of subcategories of a universal category \textbf{NCP} and the fundamental problem of quantum information theory gains more general perspective. The categorical description allows for a natural use of the notions of equivalence and representations, notions that will be exhaustively discussed elsewhere.     

This new perspective allows to introduce the notion of statistical categories and groupoids again as a natural extension of the notion of statistical manifold.    In doing so, a generalised Cramer-Rao inequality can be readily obtained and a notion of the Fisher-Rao metric which is adapted to this context and that brings a natural connection with the recent ideas in the geometry of the theory of information involving groupoids takes a new perspective.    Statistical categories provide parametric models of both the states and the channels in a coherent way.    The problem of the uniqueness of the categorical Fisher-Rao metric thus obtained will be address in subsequent publications.
Finally, it is also remarkable that the theory of non-local games can be addressed too in the proposed formalism.  The use of the groupoidal description of quantum mechanical systems could provide new insight into the algebraic structures governing such problems.

\section*{Funding}

This work has been supported by the Madrid Government (Comunidad de Madrid-Spain) under the Multiannual Agreement with UC3M in the line of “Research Funds for Beatriz Galindo Fellowships” (C\&QIG-BG-CM-UC3M), and in the context of the V PRICIT (Regional Programme of Research and Technological Innovation).
The authors acknowledge financial support from the Spanish Ministry of Economy and Competitiveness, through the Severo Ochoa Programme for Centres of Excellence in RD (SEV-2015/0554), the MINECO research project  PID2020-117477GB-I00,  and Comunidad de Madrid project QUITEMAD++, S2018/TCS-A4342.
F.D.C. thanks the UC3M, the European Commission through the Marie Sklodowska-Curie COFUND Action (H2020-MSCA-COFUND-2017- GA 801538) and Banco Santander for their financial support through the CONEX-Plus Programme. 
G.M. would like to thank partial financial support provided by the Santander/UC3M Excellence  Chair Program 2019/2020, and he is also a member of the Gruppo Nazionale di Fisica Matematica (INDAM), Italy.

\addcontentsline{toc}{section}{References}
{\footnotesize


}

\end{document}